\documentclass{SciPost}

\hypersetup{
    colorlinks,
    linkcolor={red!50!black},
    citecolor={blue!50!black},
    urlcolor={blue!80!black}
}

\usepackage[bitstream-charter]{mathdesign}
\DeclareSymbolFont{usualmathcal}{OMS}{cmsy}{m}{n}
\DeclareSymbolFontAlphabet{\mathcal}{usualmathcal}

\fancypagestyle{SPstyle}{
\fancyhf{}
\lhead{\colorbox{scipostblue}{\bf \color{white} ~SciPost Physics Proceedings }}
\rhead{{\bf \color{scipostdeepblue} ~Submission }}

\fancyfoot[C]{\textbf{\thepage}}
}

\newcommand{\mindrt}{\ensuremath{\Delta R(J,t)}}
\newcommand{\mindrw}{\ensuremath{\Delta R(J,W)}}
\usepackage{booktabs} 

\begin{document}

\pagestyle{SPstyle}

\begin{center}{\Large \textbf{\color{scipostdeepblue}{
Machine Learning Based Top Quark and W Jet Tagging to Hadronic Four-Top Final States Induced by SM as well as BSM Processes
}}}\end{center}

\begin{center}\textbf{
Jiří Kvita\textsuperscript{1}, 
Petr Baroň\textsuperscript{1$\star$}, 
Monika Machalová\textsuperscript{2}, 
Radek Přívara\textsuperscript{1}, 
Rostislav Vodák\textsuperscript{2$\dagger$}, 
Jan~Tomeček\textsuperscript{2}
}\end{center}

\begin{center}
{\bf 1} Joint Laboratory of Optics of Palacký University Olomouc and Institute of Physics of Czech Academy of Sciences, Czech Republic
\\
{\bf 2} Department of Mathematical Analysis and Applications of Mathematics, Palacký University Olomouc, Czech Republic
\\[\baselineskip]
$\star$ \href{mailto:email1}{\small petr.baron@upol.cz}\,,\quad
$\dagger$ \href{mailto:email2}{\small rostislav.vodak@upol.cz}
\end{center}

\definecolor{palegray}{gray}{0.95}
\begin{center}
\colorbox{palegray}{
  \begin{tabular}{rr}
  \begin{minipage}{0.36\textwidth}
    \includegraphics[width=60mm,height=1.5cm]{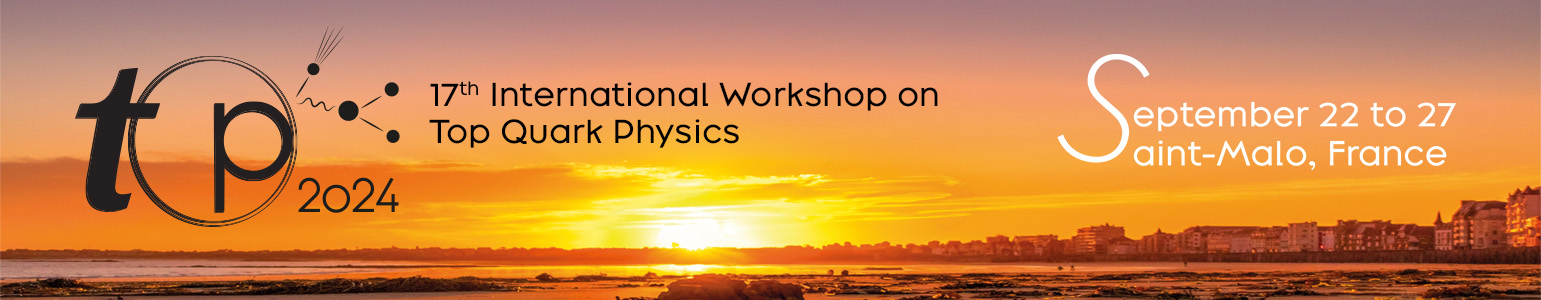}
  \end{minipage}
  &
  \begin{minipage}{0.55\textwidth}
    \begin{center} \hspace{5pt}
    {\it The 17th International Workshop on\\ Top Quark Physics (TOP2024)} \\
    {\it Saint-Malo, France, 22-27 September 2024
    }
    \doi{10.21468/SciPostPhysProc.?}\\
    \end{center}
  \end{minipage}
\end{tabular}
}
\end{center}

\section*{\color{scipostdeepblue}{Abstract}}
\textbf{\boldmath{%
We study the application of selected ML techniques to the recognition of a substructure of hadronic final states (jets) and their tagging based on their possible origin in current HEP experiments using simulated events and a parameterized detector simulation. The results are then compared with the cut-based method.
}}

\vspace{\baselineskip}

\noindent\textcolor{white!90!black}{%
\fbox{\parbox{0.975\linewidth}{%
\textcolor{white!40!black}{\begin{tabular}{lr}%
  \begin{minipage}{0.6\textwidth}%
    {\small Copyright attribution to authors. \newline
    This work is a submission to SciPost Phys. Proc. \newline
    License information to appear upon publication. \newline
    Publication information to appear upon publication.}
  \end{minipage} & \begin{minipage}{0.4\textwidth}
    {\small Received Date \newline Accepted Date \newline Published Date}%
  \end{minipage}
\end{tabular}}
}}
}




\section{Introduction}
\label{sec:intro}
Jets as hadronic final states are an inevitable consequence of the quantum chromodynamics 
(QCD)~\cite{Gross:2022hyw}, the force between strongly interacting matter constituents of quarks and gluons. 
In hadron collisions, jets are important final states and signatures of objects of high transverse momentum. 
In cases of large jet transverse momenta, i.e. with a larger Lorentz boost in the plane perpendicular to the proton beam, 
decay products of hadronically decaying $W$ bosons or top quarks are collimated so that they form one large boosted 
jet in the detector. 
This study aims to perform jet tagging for top quarks and $W$ bosons using a machine learning 
(ML) approach and compare the results with a traditional cut-based method.
%
%
%

\section{Data Samples}
\label{sec:datastructure}
Five datasets were generated using the MadGraph5 with 
different transverse momentum selection criteria on the jets and mass of hypothetical $Z'$ particle which decays into top quarks. 
Subsequently two new datasets were derived by unification of the IDs 3 and 4 (\emph{zp-sets}) 
and IDs 0--2 (\emph{pp-sets}), see Table~\ref{table}. 
\begin{table}[h!]
  \centering
  \begin{tabular}{ccc}
  \hline 
  {\bf ID} & {\bf File name} & {\bf Number of jets}\\
  \hline
  0 & ascii\_run\_XY\_pp\_2tj\_allhad\_NLO\_ptj1j2min200... & 797 363 \\
  1 & ascii\_run\_XY\_pp\_2tj\_allhad\_NLO\_ptj1j2min60...  & 446 838 \\
  2 & ascii\_run\_XY\_pp\_2tj\_allhad\_NLO\_ptj1min200... & 781 675 \\
  3 & ascii\_run\_XY\_zp\_ttbarj\_allhad\_1000GeV...                 & 449 606 \\
  4 & ascii\_run\_XY\_zp\_ttbarj\_allhad\_1250GeV...               & 388 593 \\
  \hline
  \end{tabular}
  \\[12pt]
  \hspace{6pt}
  $\rightarrow$
  \hspace{6pt}
  \begin{tabular}{ccc}
  \hline 
  {\bf ID} & {\bf File name} & {\bf Number of jets}\\
  \hline
  0 & data\_zp & 838 199 \\
  1 & data\_pp & 2 025 876 \\
  \hline
  \end{tabular}
  \captionof{table}{Table of datasets generated using MadGraph5.}
  \label{table}
  \end{table}

  \begin{figure}[h!]
  \includegraphics[width=0.45\linewidth]{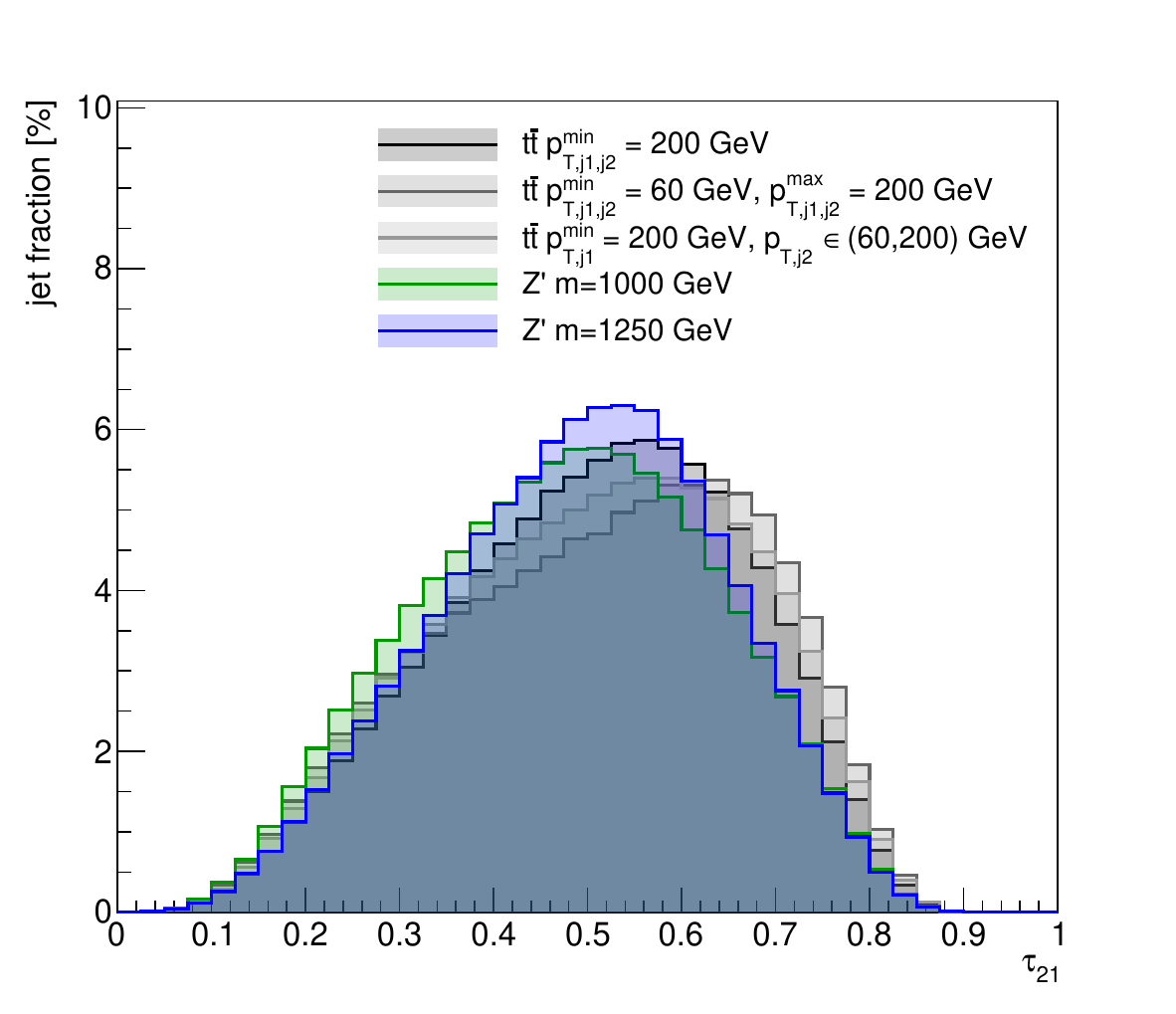}
  \hspace{0.5cm}
  \includegraphics[width=0.45\linewidth]{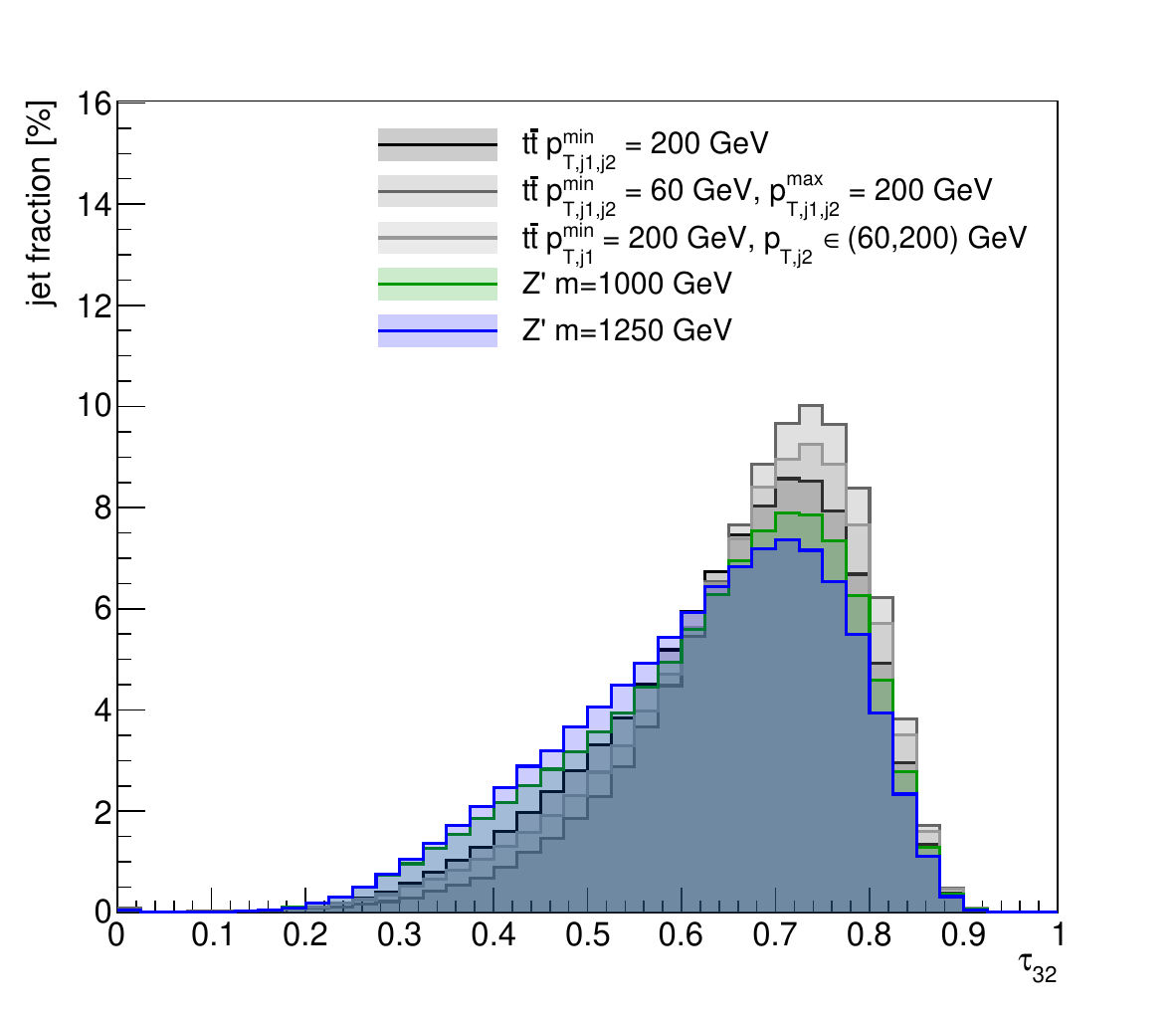}
  \label{data_zp}
  \centering
  \includegraphics[width=0.45\linewidth]{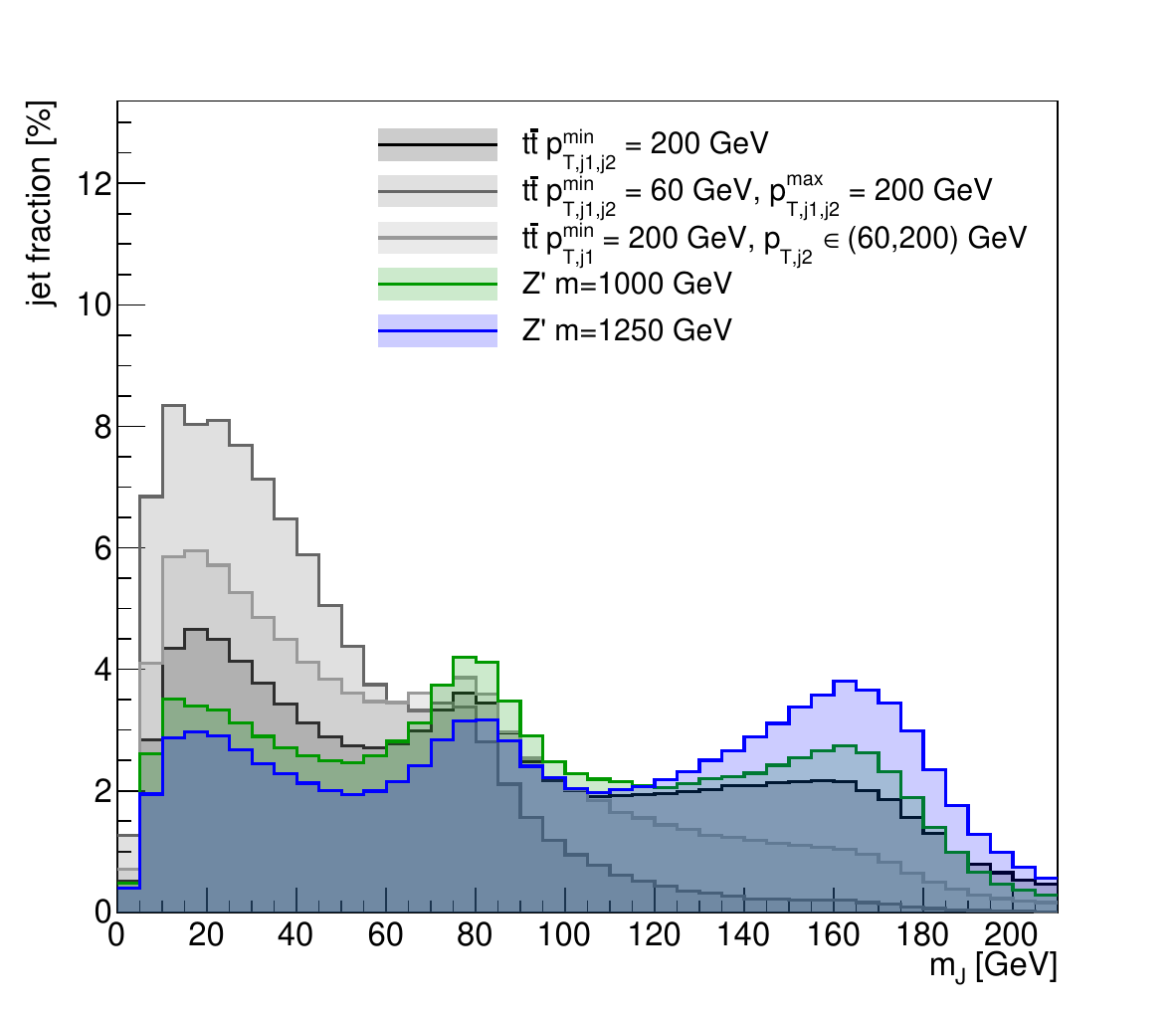}
  \hspace{0.7cm}
  \includegraphics[width=0.4\linewidth]{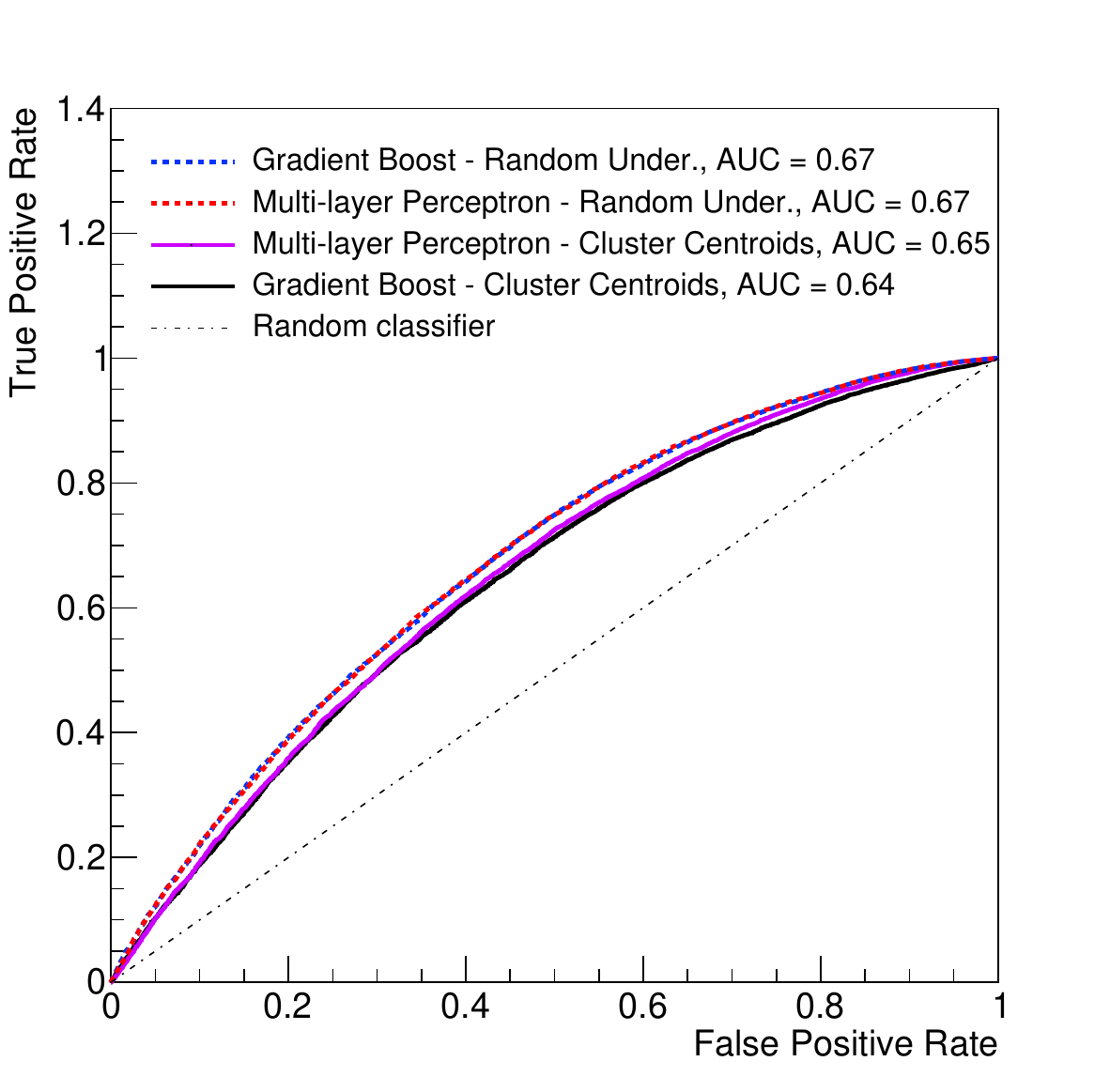}
  \caption{Shapes of the $\tau_{21}$, $\tau_{32}$ subjettiness variables (top) and the large-$R$ jet mass (bottom left) in the five samples used in training and testing of the tagging algorithms. 
  Performance of top-tagging for different classifiers shown via ROC curve (bottom right).}
  \label{data_pp}
  \end{figure}
  
  The ratios between t-jets (W-jets) and light-jets are summarized in the following tables

\begin{table}[h!]
  \centering
\begin{tabular}{ccc}
\hline
{\bf Data set} & {\bf t-jets} & {\bf light-jets}\\
\hline
\texttt{data\_zp\_t} & 86\%& 14\%\\
\texttt{data\_pp\_t} & 72.5\% & 27.5\%\\
\hline
\end{tabular}
\hspace{24pt}
\begin{tabular}{ccc}
\hline
{\bf Data set} & {\bf W-jets} & {\bf light-jets}\\
\hline
\texttt{data\_zp\_w} & 48\% & 52\%\\
\texttt{data\_pp\_w} & 42\% & 58\%\\
\hline
\end{tabular}\\[12pt]
\begin{tabular}{ccc}
  \hline
  {\bf Data set} & {\bf t-jets} & {\bf light-jets}\\
  \hline
  \texttt{data\_zp\_t} & 129 282& 21 555\\
  \texttt{data\_pp\_t} & 127 029 & 48 000\\
  \hline
  \end{tabular}
  \hspace{12pt}
  \begin{tabular}{ccc}
  \hline
  {\bf Data set} & {\bf W-jets} & {\bf light-jets}\\
  \hline
  \texttt{data\_zp\_w} & 110 735 & 121 658\\
  \texttt{data\_pp\_w} & 180 169 & 251 422\\
  \hline
  \end{tabular}
  \caption{The ratios and number of jets of t-jets (W-jets) and light-jets.}
\end{table}

Variables defined and used for each jet in the classification are as follows
\begin{table}[h!]
\resizebox{1.0\textwidth}{!}{
\begin{tabular}{rrrrrrrrrl}
\toprule
event & $\Delta$ R(J,W) & $\Delta$ R(J,t) & $p_T$ & $\eta$ & $\phi$ & $\tau_{32}$ & $\tau_{21}$ & $m$ & label \\
\midrule
0 & 0.693589 & 0.280779 & 271.076000 & -0.205725 & 1.034350 & 0.641589 & 0.304973 & 70.244600 & l \\
0 & 1.152290 & 0.542026 & 161.364000 & 1.779510 & -2.046550 & 0.678087 & 0.529191 & 67.632400 & l \\
0 & 0.505954 & 0.876577 & 88.041000 & 0.431132 & 0.073586 & 0.468017 & 0.631805 & 7.432140 & l \\
1 & 0.172936 & 0.046981 & 367.557000 & -1.193480 & -1.722920 & 0.840838 & 0.283345 & 75.302100 & w \\
1 & 0.031584 & 0.143634 & 329.300000 & -0.109191 & 1.337560 & 0.618819 & 0.205733 & 75.042200 & w \\
2 & 0.143172 & 0.050171 & 501.473000 & 0.596318 & -0.276567 & 0.605931 & 0.370552 & 171.372000 & t \\
\bottomrule
\end{tabular}
}
\caption{Defined variables for each jet.}
\end{table}

\section{Jet Selection}
\label{sec:preprocessing}
The true type jets labels are then based on the following criteria
    \begin{enumerate}
      \item truth $t$-jets:  $\mindrt{}<0.1 \land 138\,\mathrm{GeV}\leq m_J \leq 208\,\mathrm{GeV}$;
      \item truth $W$-jets; $\mindrw{}<0.1 \land 60\,\mathrm{GeV}\leq m_J \leq 100\,\mathrm{GeV}$;
      \item truth light jets: otherwise.
    \end{enumerate}
As a result, we have four subsets: zp-sets and pp-sets for t~jets, and zp-sets and pp-sets for W jets
Training sets contain $80\%$ and the test sets $20\%$ of data from the original sets.
      
%
%
\section{Methods}
\label{sec:methods}
In this section two approaches machine learning and cut-based are described in more detail.

\subsection{Classifiers}
 
\begin{itemize}
\item {\it Gradient boosting classifier} (GBC)  - 
 combining multiple simple predictors (here decision trees) to create a more powerful model
 \begin{center}\vspace{1cm}
   \includegraphics[width=0.65\linewidth]{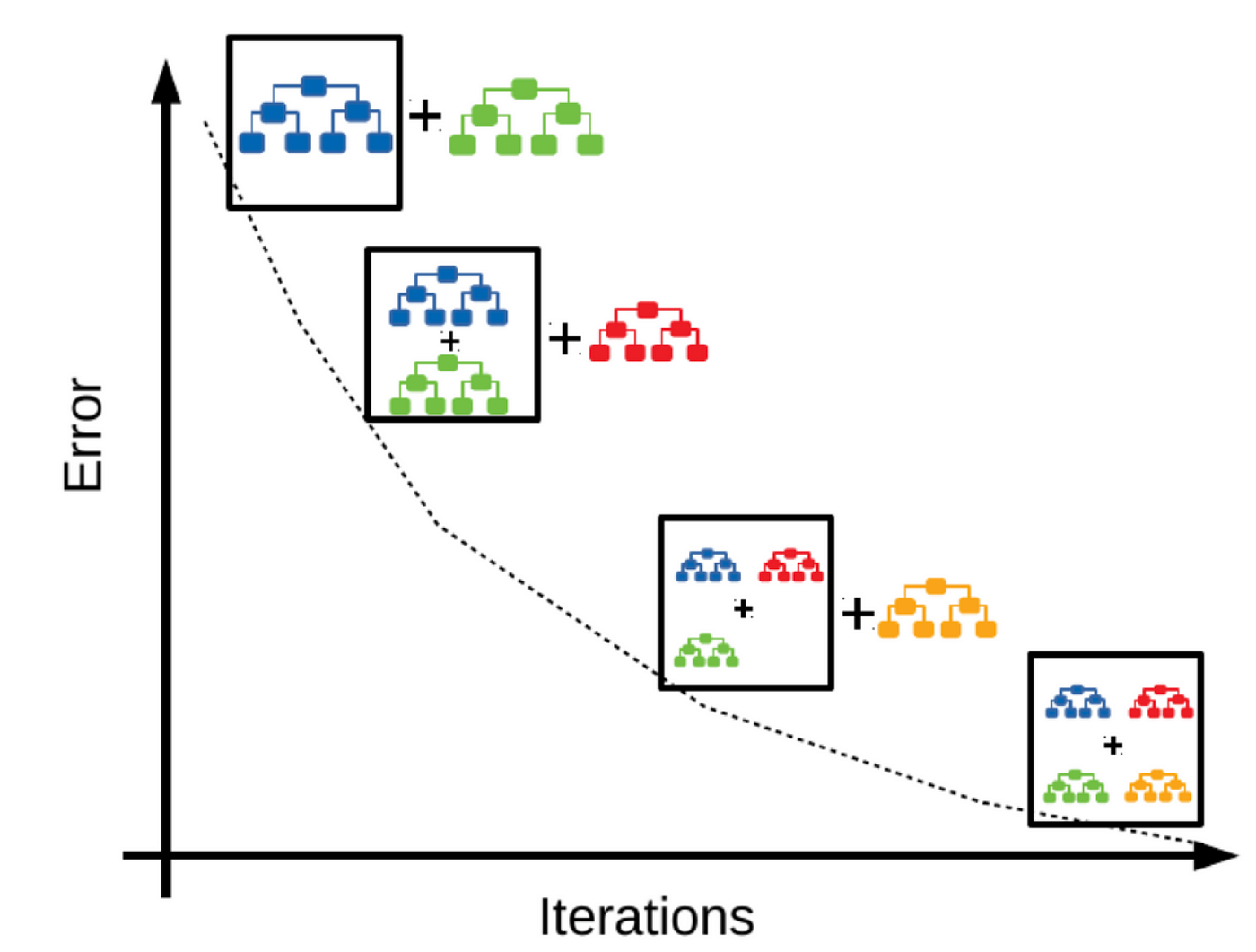}
   \captionof{figure}{GBC}
    \end{center}
\item  {\it Multi-layer Perceptron classifier} (MLP) - based on neural networks. 
\end{itemize}
\subsubsection{Undersampling} 
\begin{itemize}
\item very distorted ratio between t-jets and light-jets (in the 
direction of t-jets)
\item we settled for the undersampling applied to the training sets, which uses various techniques to remove data from the major class
\item tested undersampling techniques: {\it Random undersampling}, {\it Cluster centroids}, {\it Near miss}, {\it Repeated edited nearest neighbor}
\end{itemize}
\subsection{Cut-based algorithm}
 to identify jets coming from the hadronic decays of the $W$ boson or a top quark by a simple cut-based algorithm
\begin{itemize}
\item $W$-jets if 
{\small
$$ 0.10 < \tau_{21} < 0.60 \, \land \,   0.50 < \tau_{32} < 0.85 \, \land \, m_J \in [60, 100] \,\mathrm{GeV}$$}
\item top-jets if 
{\small
$$ 0.30 < \tau_{21} < 0.70 \, \land \,   0.30 < \tau_{32} < 0.80 \, \land \, m_J \in [138, 208] \,\mathrm{GeV}$$}
\end{itemize}

\section{Results}
\label{sec:results}
\subsection{Performance of algorithms}
\label{subsec:performanceofalgorithms}
\begin{figure}[h!]
\centering
\includegraphics[width=0.8\textwidth]{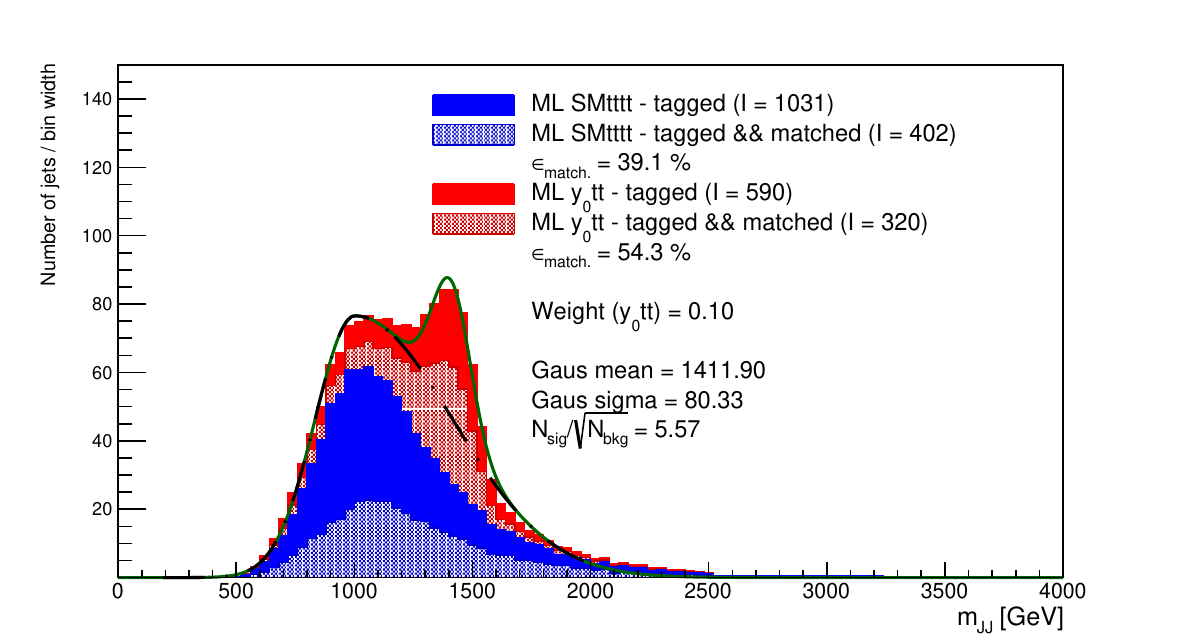}
\\
\includegraphics[width=0.8\textwidth]{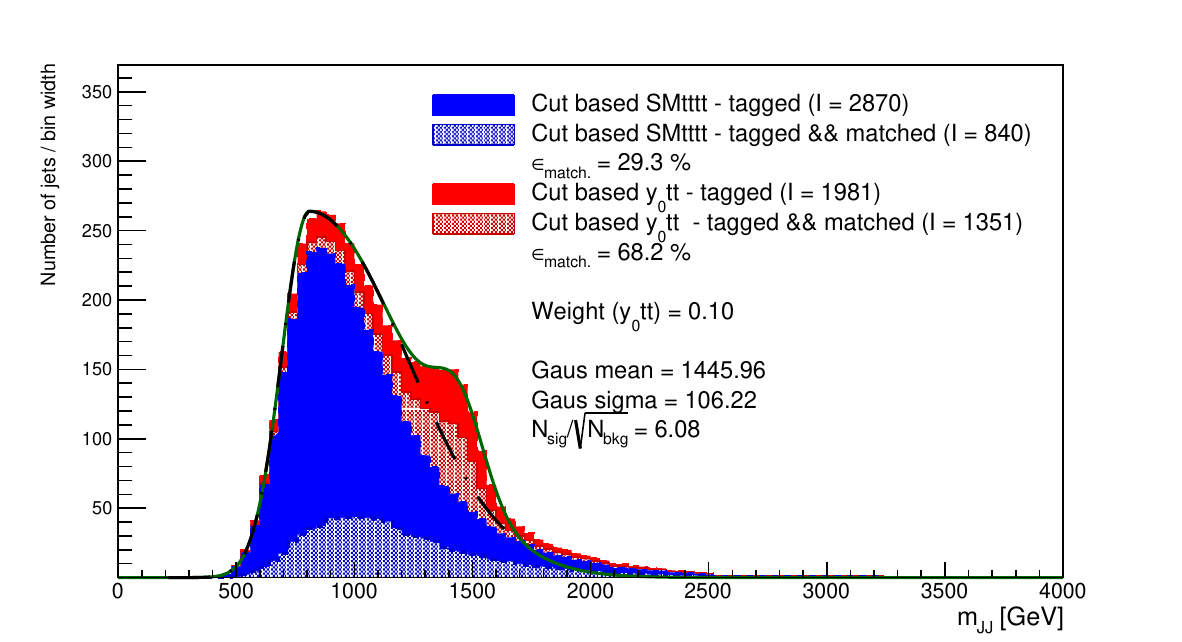}
\caption{Invariant mass of two $t$-tagged jets (top ML based, bottom cut-based algorithm) for the process of SM $t\bar{t}t\bar{t}$ (blue area) 
representing background process with the stacked signal process $t\bar{t}y_{0}\rightarrow t\bar{t}t\bar{t}$ (red area) scaled 
to its 10\%. The light red and blue areas show tagged and matched jets to highlight the tagging efficiencies. The background 
fit is given by black line using Bifurcated Gaussian and green line is the Gaussian signal fit.}
\label{peaks}
\end{figure}

The ML-based method performance is shown in Figure~\ref{peaks} left, while cut-based method on the right.\par
We perform an exercise of finding a signal peak over a falling background by performing a background fit 
using a Bifurcated Gaussian function and an additional Gaussian function for the the signal peak modelling. 
The signal significance calculated based on the fitted areas turns out to be slightly higher for cut-based 
method ($\mathrm{N}_{\mathrm{sig}}/\sqrt{\mathrm{N}_{\mathrm{bkg}}} \doteq 6.1$) compare to ML-based 
method ($\mathrm{N}_{\mathrm{sig}}/\sqrt{\mathrm{N}_{\mathrm{bkg}}} \doteq 5.6$).
On the other hand the signal peak mass resolution (standard deviation of signal Gaussian fit) 
is smaller in case of the ML-based method, $\sigma \doteq 80$~GeV compare the the cut-based method, $\sigma \doteq 106$~GeV.

\subsection{Comparison of best ML method and cut-based algorithm}
\label{subsec:Comparison}
\begin{figure}[h!]
\centering
\includegraphics[width=0.7\linewidth]{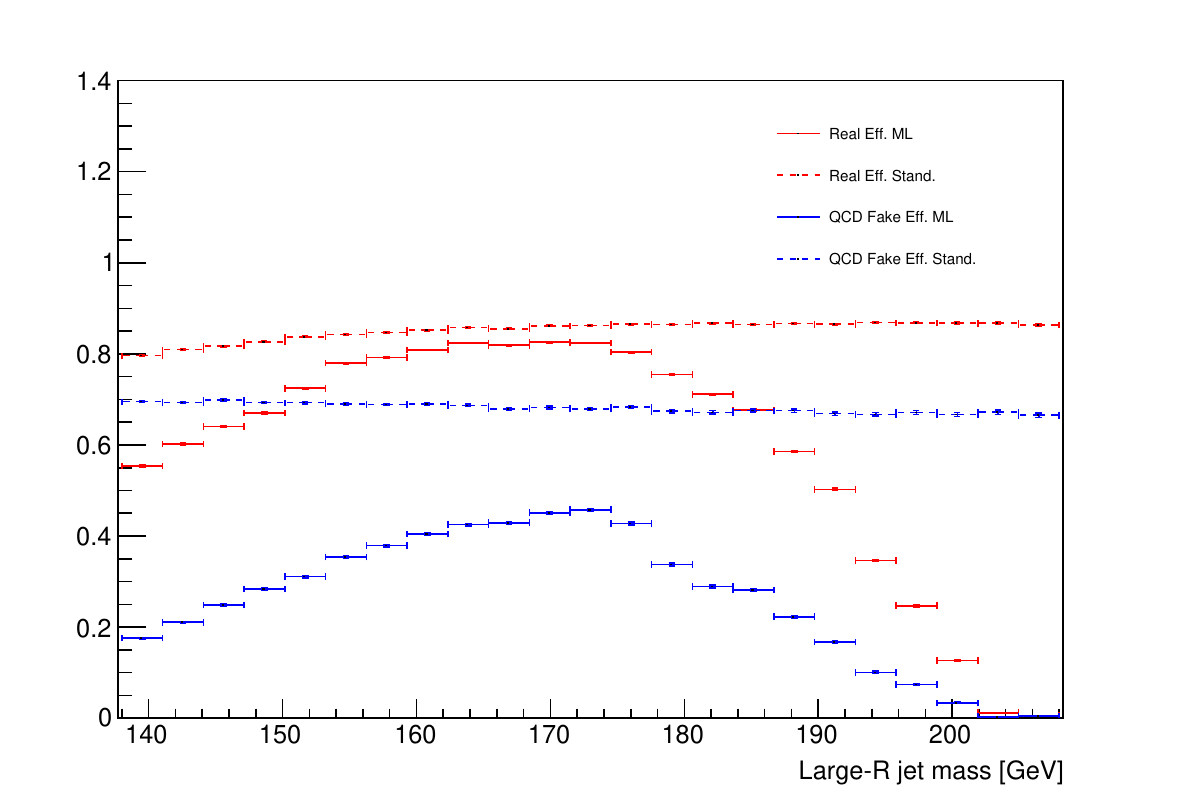}
\caption{Efficiencies using cut-based and ML, $t\bar{t}y_{0}\rightarrow t\bar{t}t\bar{t}$.}
\label{comp}
\end{figure}

In the Figure \ref{comp} we can see top tagging real efficiencies (red) and mistagging 
rates (blue) using cut-based (dashed lines) and ML-based (solid lines) of 
BSM $t\bar{t}y_{0}\rightarrow t\bar{t}t\bar{t}$ as a function of jet mass (right). 
We can see that ML based algorithms give the same real efficiencies as cut-based, 
but significantly less fake efficiencies. Where real and fake efficiencies are defined as
\begin{equation}
  \epsilon_{\mathrm{real}} = \frac{\mathrm{N(tagged~\&~matched)}}{\mathrm{N(tagged~\&~matched)}+\mathrm{N(not-tagged~\&~matched)}}
  \label{eq:realeff}
\end{equation}
\begin{equation}
  \epsilon_{\mathrm{fake}} = \frac{\mathrm{N(tagged~\&~not-matched)}}{\mathrm{N(tagged~\&~not-matched)}+\mathrm{N(not-tagged~\&~not-matched)}}
  \label{eq:fakeeff}
\end{equation}

\section{Conclusion}
The real efficiencies of cut-based method in both t-jets and $W$-jets tagging are high about 80\%, mostly flat, but unfortunatelly also having high 
mistagging rates about 65-70\%. While ML-based method has lower efficiencies, the mistagging rates are suppresed compared to cut-based method.

\section*{Acknowledgements}
The author would like to thank the grants of MSMT, Czech~Republic, GAČR 23-07110S for the support.

\end{document}